\documentclass[10pt,journal,letterpaper]{IEEEtran}

\ifCLASSINFOpdf
\usepackage{graphicx}
\usepackage{mathrsfs}
\usepackage{color}
\usepackage{amsthm}
\usepackage{amsmath}
\usepackage[square, comma, sort&compress, numbers]{natbib}
\usepackage{algorithm}
\usepackage{algorithmicx}
\usepackage{algpseudocode}

\usepackage{enumerate}
\usepackage[table,xcdraw]{xcolor}

\usepackage{marvosym}

\usepackage[square,numbers,sort&compress]{natbib}

\usepackage{array}
\newcommand{\PreserveBackslash}[1]{\let\temp=\\#1\let\\=\temp}
\newcolumntype{C}[1]{>{\PreserveBackslash\centering}p{#1}}
\newcolumntype{R}[1]{>{\PreserveBackslash\raggedleft}p{#1}}
\newcolumntype{L}[1]{>{\PreserveBackslash\raggedright}p{#1}}

\usepackage{marvosym}
\usepackage{CJK}
\usepackage{amsfonts}
\usepackage{subfigure}
\usepackage{soul}

\usepackage{multirow}

\else
\fi

\usepackage{ifpdf}
\ifpdf

\begin{document}
\title{Bluetooth-based COVID-19 Proximity Tracing Proposals: An Overview}

\author{Meng Shen,~\IEEEmembership{Member,~IEEE,}
	Yaqian Wei,
	and Tong Li*
\IEEEcompsocitemizethanks{	
\IEEEcompsocthanksitem M. Shen and Y. Wei are with the School of Computer Science, Beijing Institute of Technology, Beijing 100081, China (e-mail: shenmeng@bit.edu.cn, weiyaqianbit@foxmail.com).
\IEEEcompsocthanksitem T. Li is with 2012 Labs, Huawei. Shenzhen, 518129,  China (e-mail: li.tong@huawei.com).
\IEEEcompsocthanksitem T. Li is the corresponding author (e-mail: li.tong@huawei.com). 
}
}

\maketitle

\begin{abstract} 

Large-scale COVID-19 infections have occurred worldwide, which has caused tremendous impact on the economy and people's lives. The traditional method for tracing contagious virus, for example, determining the infection chain according to the memory of infected people, has many drawbacks. With the continuous spread of the pandemic, many countries or organizations have started to study how to use mobile devices to trace COVID-19, aiming to help people automatically record information about incidents with infected people through technologies, reducing the manpower required to determine the infection chain and alerting people at risk of infection. This article gives an overview on various Bluetooth-based COVID-19 proximity tracing proposals including centralized and decentralized proposals. We discussed the basic workflow and the differences between them before providing a survey of five typical proposals with explanations of their design features and benefits. Then, we summarized eight security and privacy design goals for Bluetooth-based COVID-19 proximity tracing proposals and applied them to analyze the five proposals. Finally, open problems and future directions are discussed.

\end{abstract}


\IEEEpeerreviewmaketitle

\section{Introduction}\label{sec:Introduction}
\IEEEPARstart{T}{he}
COronaVIrus Disease of 2019, referred to as COVID-19, has become a global pandemic and caused tens of millions of infected people and hundreds of thousands of death.
The large-scale virus infection has caused tremendous impact on people's livelihood and the economy of many countries. Many countries have to shut down cities to restrain the development of the pandemic and prevent people from working and traveling. Therefore, how to effectively curb the spread of COVID-19 has become one of the focuses of researches.  

Traditionally, to trace people who may be at risk of infection, the infected person needs to actively recall where they have been and who they have contacted during the infection period. Experts trace the people at risk of infection by constructing a relationship network and isolate them to cut off the source of infection. However, relying on the memory of the infected person is likely to miss key information. When the infected person went to a place where there were lots of people gathered, he/she could not enumerate those strangers who had come into close contact with him/her, which made it difficult for experts to analyze.

After the COVID-19 outbreak, many countries or organizations have begun to study the use of technological means to trace people who may be infected and deployed applications accordingly. These applications are expected to reduce the labor required to determine infection chains and improve the accuracy of tracing virus infections. There are already dozens of COVID-19 tracing applications. Due to the inevitable need to collect certain user information, how to protect their security and privacy has become the focus of researchers. The tracing applications can be divided into three categories based on the data collected: location data, proximity data and mixed data that includes the former two. Location data can be obtained by using Global Positioning System (GPS) to identify user's latitude and longitude, while proximity data can be obtained by using the Bluetooth function on the mobile device. Bluetooth classifies close contacts with a significantly lower false positive rate than GPS, especially in indoor environments, and it consumes lower battery \cite{bay2020bluetrace}. These Bluetooth-based applications are basically created based on five Bluetooth-based COVID-19 proximity tracing proposals.

In this article, we focus on the Bluetooth-based COVID-19 proximity tracing proposals, which can be divided into centralized proposals and decentralized proposals. Firstly, we summarized the basic workflows of the two categories of proposals and the differences between them. Then we specifically analyzed two decentralized and three centralized proposals' generation algorithms of anonymous IDs, locally stored data, uploaded data and so on. Moreover, we summarized eight security and privacy design goals of proximity tracing proposals and analyzed the five proposals according to them. We found that none of them has achieved the goals. Finally, we shed light on open problems and opportunities of Bluetooth-based COVID-19 proximity tracing proposals.

\section{Overview of Bluetooth-based Proximity Tracing Proposals}\label{sec:Status of Traffic Analysis}

With the continuous spread of the COVID-19 all over the world, many countries or organizations have successively announced Bluetooth-based proximity tracing proposals. The following are five typical proposals. In Asia, Singapore announced a privacy preserving protocol called BlueTrace \cite{bay2020bluetrace}. In Europe, the Pan-European Privacy Preserving Proximity Tracing project, referred to as PEPP-PT \cite{PEPPPT}, comprises more than 130 members across eight European countries. France’s Inria and Germany’s Fraunhofer, as members of PEPP-PT, shared a ROBust and privacy-presERving proximity Tracing protocol, referred to as ROBERT \cite{castelluccia2020robert}.  The Decentralised Privacy- Preserving Proximity Tracing proposal, referred to as DP-3T \cite{troncoso2020decentralized}, is an open protocol that ensures personal data and computation stay entirely on an individual’s phone, and this proposal was produced by a team of members from across Europe. In North America, under the influence of DP-3T \cite{troncoso2020decentralized}, Google and Apple announced a two-phase exposure notification solution, referred to as GAEN \cite{GAEN}. In the first phase, they released Application Programming Interfaces (APIs) that allow applications from health authorities to work across Android and iOS devices. In the second phase, this capability will be introduced at the operating system level to help ensure broad adoption \cite{ENFAQ}. Many applications are created based on these five proposals. Based on BlueTrace, Singapore deployed the application called TraceTogether, which is the world's first Bluetooth-based proximity tracing system deployed nationwide. The COVIDSafe application was also created based on BlueTrace and announced by the Australian Government. PEPP-PT has been implemented in Germany and they deployed the application called NTK. The French government has deployed the StopCovid application based on ROBERT to trace COVID-19. Ketju based on DP-3T was trialed in Finland and it's among the first to use a decentralised approach to proximity tracing based on DP-3T in Europe. Many countries have released open source applications based on GAEN, such as Corona-Warn-App \cite{CWA} in Germany, Stopp Corona in Austria, SwissCovid in Switzerland, Immuni in Italy and COVID Tracker in Ireland.

\subsection{Centralized and Decentralized Proposals}
\begin{figure}[h]
	
	\begin{minipage}{0.49\textwidth}
		\centering
		\includegraphics[height=11cm]{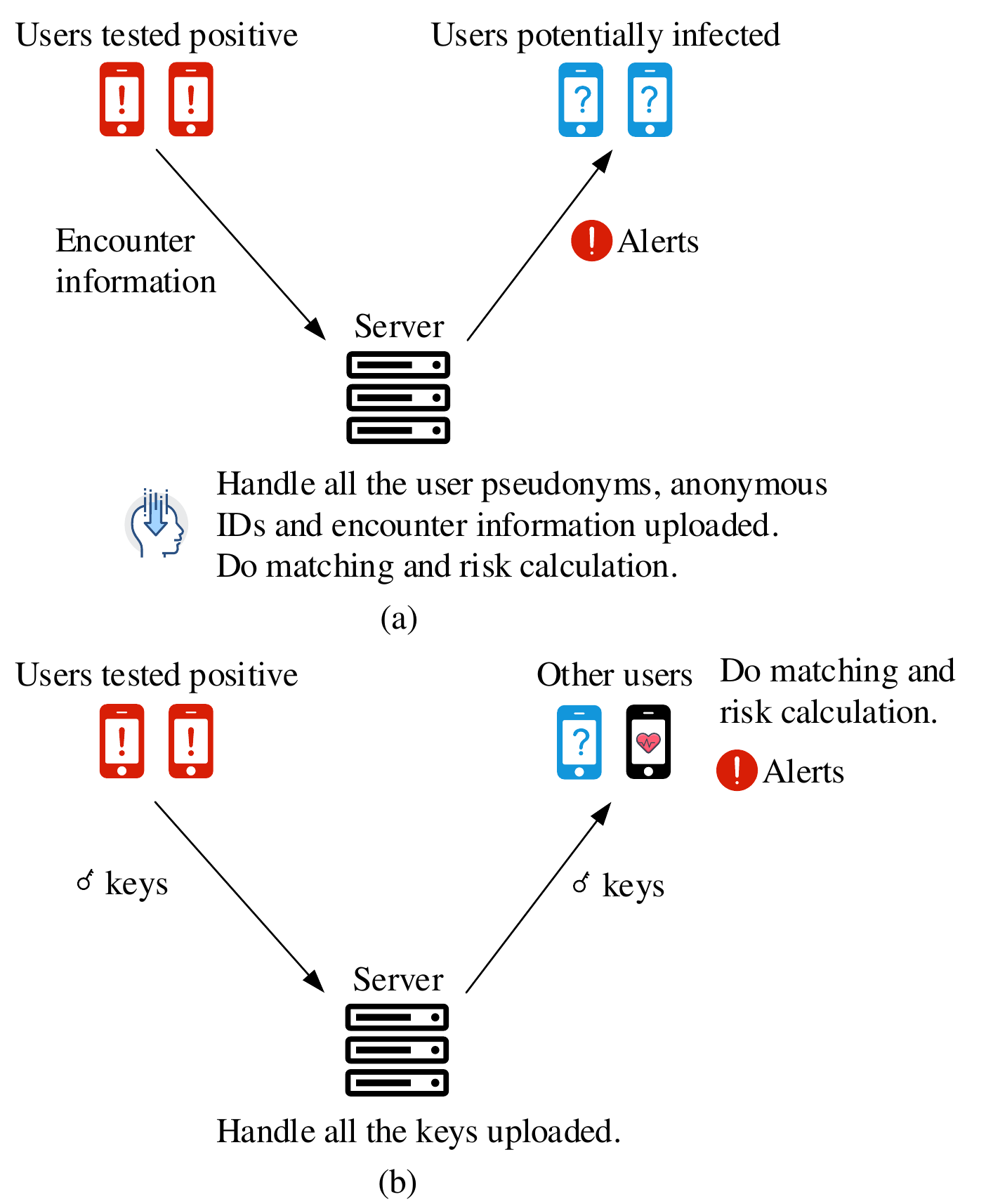}
	\end{minipage}
	\caption{Workflows of Centralized and Decentralized Proposals: (a) Centralized; (b) Decentralized. }\label{fig:centralized_and_decentralized}
\end{figure}

According to the role of the server in the proximity tracing proposals, Bluetooth-based proximity tracing proposals can be divided into two categories. One is centralized proximity tracing proposals, such as BlueTrace of Singapore, PEPP-PT of Europe and ROBERT of France. The other is decentralized proximity tracing proposals, such as GAEN and DP-3T of Europe. Figure \ref{fig:centralized_and_decentralized} (a) (b) shows the workflows of centralized and decentralized proposals, respectively. 

In the centralized proximity tracing proposals, users broadcast and receive encounter information (anonymous ID, transmission time, etc.) via Bluetooth. When users are infected with COVID-19, they can upload the encounter information to a central server, which analyzes the encounter information and determines whether any related user is at risk of infection and notifies them. The server plays a vital role in the workflow of centralized proposals and can handle the encounter information between users and analyze it.

In the decentralized proximity tracing proposals, when users are infected with COVID-19, the keys related to the generation of anonymous IDs is uploaded to the server. Then the server simply passes the keys of these positive users to other users, who regenerate anonymous IDs and analyze whether they are at risk of infection. The server only plays the role of storing and distributing keys uploaded.

The difference analysis between centralized and decentralized proposals is shown in Table \ref{tab: centralized_and_decentralized}. The back-end server using the centralized proposals handles each user's pseudonym (unique pseudo-random identifier) and encounter information. The weakness is that it can associate all the anonymous IDs of each user with his pseudonym. This allows operators of back-end servers to monitor user's behaviors. The centralized tracing proposals have been strongly criticized by privacy advocates and other stakeholders in the technical community, who believe that the centralized tracing proposals provide the government with information that can be used to reverse-engineer personal information about individuals \cite{GSMA}. Singapore and Italy have stated that they will switch from centralized applications to decentralized applications. The issue of trust has also prompted the German government favoring a centralized proposal before to adopt a decentralized one. The French Parliament debated similar concerns.

\begin{table*}[t]
	\renewcommand{\arraystretch}{1.3}
	\caption{The Differences Between Centralized and Decentralized Proposals } \label{tab: centralized_and_decentralized}
	\centering
	\resizebox{\textwidth}{!}{
		\begin{tabular}{|p{3cm}|p{6cm}|p{6cm}|}
			
			\hline
			\multicolumn{1}{|c|}{\textbf{}}                       & \multicolumn{1}{c|}{\textbf{Centralized}}       & \multicolumn{1}{c|}{\textbf{Decentralized}} \\ \hline
			Information obtained by the server                                   &All the user pseudonyms, anonymous IDs and encounter information uploaded by the users tested positive                   & All the keys uploaded by users tested positive                  \\ \hline
			The role that the server plays                                       &Analyze the information obtained and determine whether the related users may be at risk of infection                  & Store and distribute the keys                  \\ \hline
			The data volume communicated between the mobile device and the server                                       & The data uploaded by the users tested positive is small                   & The server needs to periodically distribute keys to all the users, which means the data volume greatly exceeds that of the centralized proposals                  \\ \hline
			
	\end{tabular}}
	
\end{table*}

\subsection{Comparison Between Proximity Tracing Proposals}
In this section, the two decentralized proximity tracing proposals (Table \ref{tab: decentralized} introduces GAEN and three designs of DP-3T) and the three centralized proximity tracing proposals (Table \ref{tab: centralized} introduces BlueTrace, PEPP-PT and ROBERT) are analyzed.


The two decentralized proposals have roughly similar processes, and the specific difference is reflected in the different algorithms for generating anonymous IDs. In the low-cost design, the seed keys of one user are linkable. In the formula, H represents the hash function and t represents the current day, the seed key of which can be hashed to generate that of the next day. Thus only the seed key of the first day is needed to generate all the anonymous IDs for the next few days. AnonoID represents an anonymous ID. PRF is a pseudo-random function. PRG is a pseudorandom generator. Str is a fixed, public string. Each seed key can be used to generate all the anonymous IDs of the day. In the formula of the unlinkable design, Epochs i are encoded relative to a fixed starting point shared by all the entities. LEFT128 takes the leftmost 128 bits of the hash output. This design generates a seed key for each epoch i and hashes it to generate anonymous IDs. Thus, all of these seek keys are unlinkable. The user can choose the time period for uploading, then the server regenerates these hash values based on the seed key uploaded by the user and puts them into a cuckoo filter before sending them to other users. Compared with the low-cost design, the unlinkable design provides better privacy attributes with increased bandwidth. The hybrid design uses a time window w, whose length is an integer multiple of the anonymous ID's valid period, to reduce the valid period of a seed key. The user can also select the time period and time window w for uploading. Compared with low-cost designs, this design requires more bandwidth and storage space, but less than that of the unlinkable design. GAEN is similar to the hybrid design of DP-3T. It corresponds to the case where the time window of the hybrid design is one day but has been upgraded in the generation algorithm of anonymous ID. In the formula, SecSeed represents secondary seek key and PriSeed represents primary seek key. HKDF is a key derivation algorithm. It first generates a primary seed key every day that is unassociated with each other. Then it uses the primary seed key to generate a secondary seed key, which is used to generate an anonymous ID.

The servers in the three centralized proposals all grasp user pseudonyms, anonymous IDs and encounter information uploaded. BlueTrace needs to collect user's phone number and associate the number with user's pseudonym. Their differences are also mainly reflected in the different algorithms for generating anonymous IDs. In addition to generating an anonymous ID using a key known only to itself, the server in ROBERT also uses the anonymous ID and a key known only to itself to generate encrypted country code to implement the proposal across the country. In these proposals, the server plays an important role. 

In the decentralized proposals, the users tested positive store the encounter information broadcast by other users on mobile device, and upload the keys that generate the anonymous IDs. While in the centralized proposals, the users tested positive store and upload the encounter information broadcast by other users. The server handles users’ pseudonyms and anonymous IDs. So as long as a user uploads the encounter information, the server can infer whether there are related users at risk of infection. In the decentralized proposals, because the server is responsible for storing and distributing data, users need to upload keys so that other users can acquire keys and regenerate anonymous IDs to match.

DP-3T proposes three different decentralized designs with different bandwidth and privacy requirements. The low-cost design requires the minimum bandwidth and provides the weakest privacy. The unlinkable design requires the maximum bandwidth and provides the strongest privacy. And the bandwidth and privacy of the hybrid design is between the low-cost design and the unlinkable design. GAEN is similar to the hybrid design of DP-3T, but it has a better anonymous ID generation algorithm. In the above three centralized proposals, the information collected from users and the anonymous ID generation algorithm are different. BlueTrace needs to collect the user's phone number, while ROBERT does not need. ROBERT uses a secret key known only to the server to encrypt the country/region code as part of encounter information, while the other two proposals do not. In decentralized proposals, a user uploads keys related to his/her own anonymous IDs to the server. But in centralized proposals, the user uploads the encounter information related to other users' anonymous IDs to the server.

\begin{table*}[t]
	\normalsize
	\renewcommand{\arraystretch}{1.3}
	\caption{Comparison of The Two Decentralized Proximity Tracing Proposals } \label{tab: decentralized}
	\centering
	\resizebox{\textwidth}{!}{
		\begin{tabular}{|p{3cm}|p{3cm}|p{4.5cm}|p{4cm}|p{5cm}|}
			
			\hline
			\multirow{3}*{} & \multicolumn{4}{c|}{Decentralized} \\
			\cline{2-5}
			& \multirow{2}*{GAEN} & \multicolumn{3}{c|}{DP-3T} \\
			\cline{3-5}
			&  &  1. Low-cost design & 2. Unlinkable design & 3. Hybrid design  \\ \hline
			Personal information                                       & None                    &    \multicolumn{3}{c|}{None}                 \\ \hline
			User pseudonym                                      & None                    &    \multicolumn{3}{c|}{None}                 \\ \hline
			Generation algorithm of anonymous ID                                     &$SecSeed_t = HKDF(PriSeed_t)$ $AnonID_i = PRG(SecSeed_t, i)$                  
			&$Seed_t = H(Seed_{t-1})$ $AnonID_1||\cdots||AnonID_n = PRG(PRF(Seed_t, str))$       &$AnonID_i= LEFT128(H(Seed_i))$     &$AnonID_{w,1}||\cdots||AnonID_{w,n}=PRG(PRF(Seed_w, str))$
			\\ \hline
			Generating location                                       & Mobile device                    &    \multicolumn{3}{c|}{Mobile device}                 \\ \hline
			Identity of infected users                                      & The server does not know                    &    \multicolumn{3}{c|}{The server does not know}                  \\ \hline
			Data saved on mobile devices                                       & Anonymous ID and Associated Encrypted Metadata (version and transmit power level)                  & Anonymous ID , exposure measurement (for example, signal attenuation) and receiving date         &Hash string of anonymous ID and time, exposure measurement and receiving date       &Anonymous ID, exposure measurement, time window for receiving anonymous ID                 \\    \hline
			Data uploaded                                       & $(PriSeed_t,t)$                  &$(Seed_t,t)$          &$(Seed_i,i)$ in the time period selected     &$(Seed_w,w)$ in the time period selected                 \\  \hline
	\end{tabular}}
	
\end{table*}

\section{Security and Privacy Analysis of Bluetooth-based Proximity Tracing Proposals}\label{sec:feature}

\subsection{Security Analysis}

This section summarizes the security design goals required for the Bluetooth-based proximity tracing proposals based on six types of threats proposed in the STRIDE Threat Model of Microsoft \cite{shostack2014threat} and analyzes the security of five proposals. 

\subsubsection{Security Design Goals} Eight security design goals are as follows.

\textbf{Information confidentiality.} Attackers cannot obtain information transmitted by users through wireless communication.

\textbf{Information integrity.} When transmitting and storing encounter information, these proposals should ensure that the information is not tampered by unauthorized entities or can be discovered afterwards.

\textbf{Normal reception.} A user can normally receive the information broadcast by the other users after granting the application permission.

\textbf{Processing of big data.} The application works normally when it receives a large amount of encounter information.

\textbf{Avoidance of false contact.} Only when two users have close contact can they receive the information broadcast by each other.

\textbf{Real identity.} An attacker cannot claim to be a certain user.

\textbf{Authorization.} The user tested positive needs authorization or identity verification before uploading data to the server.

\textbf{Non-repudiation.} Users cannot deny that they have had close contact with someone.

\subsubsection{Security Analysis of Proposals} The analysis of these five proposals' achievement of the security design goals is as follows.

\begin{table*}[t]
	\renewcommand{\arraystretch}{1.3}
	\caption{Comparison of The Three Centralized Proximity Tracing Proposals } \label{tab: centralized}
	\centering
	\resizebox{\textwidth}{!}{
		\begin{tabular}{|p{3cm}|p{4cm}|p{4cm}|p{4cm}|p{4cm}|}
			
			\hline
			\multirow{2}*{} & \multicolumn{3}{c|}{Centralized} \\
			\cline{2-4}
			&  BlueTrace & PEPP-PT & ROBERT \\ \hline
			Personal information                                       & Phone number                    &    Not clear  &None                \\ \hline
			User pseudonym                                      & A unique random identity                    &    A unique random identity   &A unique random identity                 \\ \hline
			Generation algorithm of anonymous ID                                       & Use seed key only known to sever to encrypt user pseudonym, creation time and expiration time                   &Use seed key periodically generated by server to encrypt user pseudonym       &Use seed key only known to sever to encrypt user pseudonym and the epoch                      \\ \hline
			Generating location                                       & The server                    &  The server  &The server               \\ \hline
			Identity of infected users                                     & The server knows                   &    Not clear  &The server knows user pseudonyms but cannot link to user’s identity                \\ \hline
			Data saved on mobile devices                                       & Anonymous ID, Received Signal Strength Indication (RSSI), Device Model, timestamp, etc.                  &Anonymous ID, metadata and optional device information and device status (RSSI and TX/RX power), timestamp and optional further data (such as WiFi status)         &Encrypted country code, anonymous ID, timestamp, Message Authentication Code (MAC) and transmission time                   \\ \hline
			Data uploaded                                        & Data saved on mobile devices                 & Data saved on mobile devices         &Data saved on mobile devices             \\ \hline
	\end{tabular}}
	
\end{table*}

\textbf{Information confidentiality.} In all the proposals, a user broadcasts information to the other nearby users via Bluetooth. In this process, attackers can use tools, such as sniffer, to obtain massages broadcast by users. But attackers cannot obtain valid information by analyzing  these messages due to using the generation algorithm of anonymous IDs. In decentralized proposals, only users who may be at risk of infection can do risk calculation. In centralized proposals, only servers can decrypt the encounter information and obtain confidential information about users. 

\textbf{Information integrity.} In the two decentralized proposals, if the seed keys uploaded by users tested positive are tampered, other users cannot regenerate real anonymous IDs based on the false keys. In GAEN, anonymous IDs and Associated Encrypted Metadata (AEM) are both encrypted. If they are tampered, the encounter information regenerated by users based on the real seed keys cannot match them. In DP-3T, if the anonymous IDs or the hash value of the anonymous IDs in the encounter information is tampered, the anonymous IDs regenerated based on the seed keys cannot match them. In PEPP-PT, the anonymous IDs in the encounter information are generated by the periodically changing seed keys. When the user pseudonym decrypted is invalid, it can be determined that the information has been tampered. In BlueTrace, there are fields for integrity checking in the anonymous IDs. In ROBERT, the Message Authentication Code (MAC) in the encounter information can be used to check the integrity.

\textbf{Normal reception.} Any proximity tracing system based on Bluetooth Low Energy (BLE) is vulnerable to active attackers. This attack may cause the normal recording of anonymous IDs to stop working, thereby preventing a user from discovering the other users. This is an inherent problem with this method.

\textbf{Processing of big data.} When an attacker sends a large amount of encounter information to a user, the user's application may occupy too much memory to store the information, which may cause the application to crash. To solve this problem, the storage capacity can be set for the encounter information, but this method will also cause the application to be unable to receive more encounter information after the encounter information fills up the memory. None of these five proposals can deal with this problem.

\textbf{Avoidance of false contact.} For all the proposals, false contact incidents cannot be completely avoided. The attacker can record the information broadcast by a user and broadcast it to victims as quickly as possible. If the user is later tested positive, the victims will mistakenly believe that they are in danger. Technically savvy attackers can use large antennas to artificially increase their broadcast range. For attackers without budget restriction, they may relay and broadcast anonymous IDs extensively to create large-scale false contact incidents. All the proposals resist these attacks to the greatest extent by limiting the validity period of anonymous IDs but it cannot solve this problem completely.

\textbf{Real identity.} All the proposals use specific encryption algorithms to prevent the attacker from deriving seed keys or user pseudonyms based on the collected anonymous IDs, so the attacker cannot pretend to be a certain user.

\textbf{Authorization.} In all the proposals, users infected with COVID-19 can upload data to the server only after being authorized by health authorities.

\textbf{Non-repudiation.} In the centralized proposals, the server handles user pseudonyms, anonymous IDs generated based on the user pseudonyms and encounter information uploaded. When two users have proximity contacts, they will send encounter information including anonymous IDs to each other. When one user is tested positive and uploads encounter information to the server, the other user cannot deny the proximity contact with him/her because the sever can get another user's pseudonym from the anonymous ID in the encounter information. In the decentralized proposals, if one user is tested positive and uploads keys to the server and another user gets the keys, regenerates and matches the anonymous IDs successfully, the user cannot deny the proximity contact with another user because the keys are only known to him/her.

Based on above analysis, we listed the achievement of the five proximity tracing proposals for eight security design goals, as shown in Table \ref{tab: security_design_goals}. It can be seen that none of the five proposals can achieve the security design goals of normal reception, processing of big data and avoidance of false contacts. And all can achieve the design goals of confidentiality, integrity, real identity, authorization and non-repudiation.

\begin{table*}[t]
	\renewcommand{\arraystretch}{1.3}
	\caption{Achievement of Security Design Goals of The Five Proposals } \label{tab: security_design_goals}
	\centering
	\setlength{\tabcolsep}{7mm}{
	\begin{tabular}{|l|l|l|l|l|l|}
		\hline
		\multicolumn{1}{|c|}{\textbf{Security Design Goals}}  &   \multicolumn{1}{c|}{\textbf{GAEN}}                       & \multicolumn{1}{c|}{\textbf{DP-3T}}       & \multicolumn{1}{c|}{\textbf{BlueTrace}} & \multicolumn{1}{c|}{\textbf{PEPP-PT}} & \multicolumn{1}{c|}{\textbf{ROBERT}} \\ \hline
		Information confidentiality                                       & ${\surd}$                   & ${\surd}$                           & ${\surd}$                                 & ${\surd}$   & ${\surd}$               \\ \hline
		Information integrity                                      & ${\surd}$                   & ${\surd}$                               & ${\surd}$                                     & ${\surd}$   & ${\surd}$                   \\ \hline
		Normal reception                                       & ${\texttimes}$                   & ${\texttimes}$                            & ${\texttimes}$                                   & ${\texttimes}$   & ${\texttimes}$                \\ \hline
		Processing of big data                                     & ${\texttimes}$                   & ${\texttimes}$                            & ${\texttimes}$                                   & ${\texttimes}$   & ${\texttimes}$                \\ \hline
		Avoidance of false contact                                     & ${\texttimes}$                   & ${\texttimes}$                            & ${\texttimes}$                                   & ${\texttimes}$   & ${\texttimes}$                \\ \hline
		Real identity                                      & ${\surd}$                   & ${\surd}$                               & ${\surd}$                                     & ${\surd}$   & ${\surd}$                   \\ \hline
		Authorization                                      & ${\surd}$                   & ${\surd}$                               & ${\surd}$                                     & ${\surd}$   & ${\surd}$                   \\ \hline
		Non-repudiation                                      & ${\surd}$                                     & ${\surd}$                              & ${\surd}$                                     & ${\surd}$   & ${\surd}$                   \\ \hline
	\end{tabular}}
	
\end{table*}

\subsection{Privacy Analysis}
This section summarizes the eight privacy design goals required for the Bluetooth-based proximity tracing proposals based on six data protection principles of General Data Protection Regulation (GDPR) of the European Union \cite{hoofnagle2019european} and analyzes the privacy of the five  proposals.

\subsubsection{Privacy Design Goals} The eight privacy design goals are as follows.

\textbf{Right of access.} Users shall have the right to obtain confirmation as to whether or not personal data concerning them are being processed, and where that is the case, access to the personal data and the following information: the purposes of the processing, the categories of personal data concerned, the period that personal data will be stored etc. 

\textbf{Data minimisation.} Adequate, relevant and limited to what is necessary in relation to the purposes for 
which they are processed. 

\textbf{Right to erasure.} Users shall have the right to obtain the erasure of personal data concerning them without undue delay and the system shall have the obligation to erase personal data without undue delay.

\textbf{Storage limitation.} Save the data for necessary limited time and then erase it.

\textbf{Untraceability.} Users' locations cannot be exposed based on the information broadcast by them.

\textbf{Protection of infected users.} The identity of infected users should not be exposed to unauthorized entities. 

\textbf{Protection of risky users.} The identity of risky users should not be exposed to unauthorized entities.

\textbf{Protection of interaction information.} The interaction information which reflects close-range physical interactions between users should not be exposed to unauthorized entities.

\subsubsection{Privacy Analysis of Proposals}The analysis of the five proposals' privacy design goals is as follows.

\textbf{Right of access.} In all proposals, the applications provide introduction to users before they use specific functions. They inform users the permission users should grant, the purpose of collecting the data, the data they will collect, the period that personal data will be stored etc.

\textbf{Data minimisation.} All of the five proposals indicated that the location information of mobile phones would not be collected. The processing amount of personal data is limited to the minimum amount of data required by the system, and no unnecessary personal data is collected

\textbf{Right to erasure.} In the five proposals, users have the right to stop using the applications and delete personal data at any time.

\textbf{Storage limitation.} All of the five proposals limit the number of days the data can be kept. Once the data expires, it will be deleted, ensuring the accuracy of the stored data.

\textbf{Untraceability.} In the two decentralized proposals, since users use encryption algorithms to generate anonymous IDs that change periodically, other entities cannot link to users by analyzing anonymous IDs they broadcast. In the three centralized proposals, other entities cannot link to user pseudonyms by analyzing anonymous IDs unless they have encryption keys. But encryption keys are only handled by servers. Consequently, other entities cannot link to users by analyzing the anonymous IDs they broadcast. 

\textbf{Protection of infected users.} The data uploaded by the infected user is not related to personal information. In the decentralized proposals, the data uploaded is keys, and in the centralized proposals, it is the encounter information. However, attackers can determine that the user is uploading a large amount of data to the server by tracing phone numbers of health authorities or observing the network traffic, inferring that the user has been tested for COVID-19 and diagnosed. These attackers can be Internet Service Providers (ISPs), network operators, or hackers who set up malicious access points or sniff public WiFi networks.

\textbf{Protection of risky users.} In the two decentralized proposals, the server sends seed keys uploaded by the infected users to the other users, who use the seed keys locally to regenerate the anonymous IDs and calculate the risk score. These seed keys are not associated with the identity of the user at risk of infection, so the decentralized proposals will not disclose information about the user at risk of infection to the others. In PEPP-PT and ROBERT, all the users will periodically request to the server to update the risk score. The format of the reply is the same regardless of whether the user has a risk of infection. Therefore, if the server is credible and the communication channel is confidential, the eavesdropper cannot distinguish which user has a risk of infection. In BlueTrace,  because the server uses user's phone number to notify them of the risk of COVID-19 infection, the information of the user with high risk may be leaked through the attacker's tracking of the health authority's phone number.

\textbf{Protection of interaction information.} In the two decentralized proposals, the system will not disclose any information about the interaction between two users to any entity. The anonymous IDs derived from the keys uploaded by an infected user has nothing to do with whoever had interacted with this user. In the three centralized proposals, only the server can learn an infected user's interaction information by analyzing the encounter information uploaded by the user. If the server is trusted, the other unauthorized parties will not learn about these interaction information.

Based on the above analysis, we listed the achievement of the five proximity tracing proposals for the eight privacy design goals, as shown in Table \ref{tab: privacy_design_goals}. It can be seen that all the proposals have achieved the same privacy in terms of right of access, data minimisation, right to erasure, storage limitation, untraceablility and protection of interaction information. And none of them achieves the privacy goal of protecting infected users. For the privacy design goal of protecting risky users, since BlueTrace needs to collect users’ phone numbers, it may disclose the information of users at risk of infection. In terms of privacy analysis, BlueTrace achieves less privacy design goals than the other proposals.

\begin{table*}[t]
	\renewcommand{\arraystretch}{1.3}
	\caption{Achievement of Privacy Design Goals of The Five Proposals } \label{tab: privacy_design_goals}
	\centering
	\setlength{\tabcolsep}{7mm}{
	\begin{tabular}{|l|l|l|l|l|l|}
		\hline
		\multicolumn{1}{|c|}{\textbf{Privacy Design Goals}}  &   \multicolumn{1}{c|}{\textbf{GAEN}}                       & \multicolumn{1}{c|}{\textbf{DP-3T}}       & \multicolumn{1}{c|}{\textbf{BlueTrace}} & \multicolumn{1}{c|}{\textbf{PEPP-PT}} & \multicolumn{1}{c|}{\textbf{ROBERT}} \\ \hline
		Right of access                                   & ${\surd}$                   & ${\surd}$                               & ${\surd}$                                     & ${\surd}$   & ${\surd}$                   \\ \hline
		Data minimisation                                      & ${\surd}$                   & ${\surd}$                               & ${\surd}$                                     & ${\surd}$   & ${\surd}$                   \\ \hline
		Right to erasure                                         & ${\surd}$                   & ${\surd}$                               & ${\surd}$                                     & ${\surd}$   & ${\surd}$                   \\ \hline
		Storage limitation                                  & ${\surd}$                   & ${\surd}$                               & ${\surd}$                                     & ${\surd}$   & ${\surd}$                   \\ \hline
		Untraceability                                    & ${\surd}$                   & ${\surd}$                               & ${\surd}$                                     & ${\surd}$   & ${\surd}$                   \\ \hline
		Protection of infected users                                     &  ${\texttimes}$                   &  ${\texttimes}$                               &  ${\texttimes}$                                    &  ${\texttimes}$   &  ${\texttimes}$                    \\ \hline
		Protection of risky users                                      & ${\surd}$                   & ${\surd}$                               & ${\texttimes}$                                    & ${\surd}$   & ${\surd}$                   \\ \hline
	    Protection of interaction information                                      & ${\surd}$              & ${\surd}$                           & ${\surd}$                                     & ${\surd}$   & ${\surd}$                   \\ \hline
	\end{tabular}}
	
\end{table*}

\section{Open Problems and Opportunities}

At present, the research on Bluetooth-based proximity tracing proposals is still in the stage of continuous exploration, and researchers face many challenges during this process.

\subsection{Precise Proximity Measurement and Risk Calculation}

Precise proximity measurement and risk calculation are the key steps in tracing COVID-19. 
GAEN standardizes four scores which are attenuationScore, daysSinceLastExposureScore, durationScore and transmissionRiskScore, and then it multiples these scores to calculate the risk value of infection. SwissCovid, which is based on GAEN, divides the attenuation into three intervals by using two attenuation values before assigning different weight values to each interval. It uses GAEN API to request user’s continuous attenuation time in different intervals and get the risk value of infection by calculating the weighted sum of continuous attenuation time. GAEN is still evolving, and the measurement and calibration between different operating systems and different mobile phone models of these parameters such as attenuation values, continuous contact time, thresholds and weights are still not completed. To accurately estimate the distance between two users, GAEN released a Bluetooth Low Energy RSSI Calibration Tool to calibrate as many devices as possible. It collects the RSSI Correction and the transmit power of different mobile phone models to improve the calculation consistency of attenuation values of all devices \cite{BLE}. GAEN currently uses this rough calibration method as a stopgap measure. The United Kingdom believes that the method of GAEN measuring distance through RSSI is inaccurate and creates its own centralized tracing application. NHSX, the digital innovation unit of the British National Health Service, released the NHS COVID-19 application \cite{NHSX} and tried it out on the Isle of Wight, but many technical challenges have been identified through system testing.

Measuring the distance between users may consider the mutual enhancement of Bluetooth and ultrasonic ranging. In Bluetooth-based proximity tracing proposals, mobile devices broadcast anonymous IDs using Bluetooth Low Energy (BLE), in which the attenuation of Bluetooth signals is generally used to indirectly represent the distance between users. In addition, ultrasound is also a way to measure distance, which is more accurate and does not depend on special hardware. In a scenario where the distance between users is greater than the officially considered safe distance, the attenuation of the Bluetooth signal can be used to represent the distance between users, because this scenario does not require an accurate distance measurement. When the users are in close contact, for example, when the two users perform handshake and other close actions, ultrasonic assisted ranging perhaps can be triggered as needed at this time to provide calibration for distance measurement.

\subsection{Security and Privacy Guarantee}

The researchers conducted experiments on the privacy and security risks of GAEN in the real world. In this experiment, the researchers proved that the current framework design is vulnerable to two kinds of attacks \cite{baumgartner2020mind}. One attack is to profile the infected person and possibly de-anonymize it. The researchers used mobile devices as a Bluetooth sniffer to capture anonymous IDs broadcast by them passing through six locations. The captured data appeared random and could not be associated with a single user. However, after a user is tested positive and continuously uploads the primary keys, the result is completely different. By generating a user’s anonymous IDs and matching with the anonymous IDs received by the Bluetooth sniffers at six locations, they can accurately know which locations the user has visited, and the user’s route map can be portrayed based on the time information. Thus, they can collect a lot of information about the user and cancel its anonymity. Because the code of GAEN is not open source and the API can only be used by health authorities, an analog tracker that conforms to the anonymous ID encryption specification in the GAEN API is used in this experiment. Another type of attack is a relay-based wormhole attack, in which an attacker constructs a fake contact event and may seriously affect a tracing application built on GAEN. The researchers built a multi-location wormhole by integrating Bluetooth Low Energy (BLE) and the Raspberry Pi. First, the worm device sends the encounter information collected from a location to the central Message Queuing Telemetry Transport (MQTT) server. The server distributes the received messages among the worm devices. These devices will copy the beacon within the validity period of the anonymous ID (10 minutes) and rebroadcast. Finally, the researchers established a logical connection between the mobile devices 40 kilometers away, but in fact they did not have real contacts. This wormhole attack budget is relatively low, and attackers can use higher-than-normal signal strength and/or high-gain antennas to significantly increase the scope of each wormhole device. Therefore, an attacker may establish false connections between a large number of users and expand the number of people who need to be tested and isolated, causing unnecessary panic. Because GAEN is unavailable, the researchers used DP-3T that inspired GAEN as an alternative. All the COVID-19 tracing applications designed based on GAEN are vulnerable to these two attacks. For the centralized proposals, since the server handles every user’s pseudonym, each user can be monitored. Thus it is necessary to ensure that the server is credible and will not disclose information.

To promote the progress of these proposals in terms of security and privacy, governments or research institutions can open-source their proposals and use everyone's power to find a better evolutionary direction.

\subsection{Interoperability of Applications} \label{sec:evaluation}

GAEN develops a Bluetooth-based proximity tracing system on Android and iOS platforms to improve the security and privacy of the Bluetooth function used in the proposal, but the framework may not be available on other platforms \cite{GSMA}. The European countries believe that Apple’s mobile phones restrict the use of Bluetooth background scanning by third-party applications. Their users’ mobile devices need to keep Bluetooth on and active at all times, which will negatively affect battery life and device availability, making their own proposals impossible and turning to GAEN to build applications. In addition, when performing accurate proximity measurements based on radio signal strength, devices with different technical characteristics need to be considered.

\subsection{Large-scale User Group}

To ensure the validity of the proximity tracing applications, a large number of users must download these applications and grant application permissions. When only a small number of people choose to use tracing applications, none of these proposals can play their true role. To protect the privacy of users, users infected with COVID-19 can decide whether to upload data to the server in the five proposals mentioned above. If only some users choose to upload data, these proposals cannot effectively trace COVID-19. The applications are unavailable in areas lacking the 3rd or 4th generation mobile communication technologies. And for the elderly, children, and people with difficult family conditions, they may not have qualified mobile devices and consequently cannot use these applications. For example, the smartphone penetration in India and Bangladesh is very low, which is 25.3\% and 18.5\%, respectively \cite{chowdhury2020covid}. Moreover, in some countries or regions that are highly concerned about personal privacy, the security and privacy risks of the applications are also a reason that prevents people from using them.

The government should increase publicity efforts for such type of applications on the basis of protecting user’s safety and privacy and try to implement this function on other portable devices to reduce the threshold for using them \cite{kleinman2020digital}.

\subsection{Powerful Infection Detection Capability}

In any country, ensuring a strong COVID-19 infection test capability is the basis for preventing the spread of COVID-19. Health authorities must be able to test accuratly whether people are infected with COVID-19 on a large scale so that these auxiliary tracing proposals can achieve their functions. If users fail to test in time and get accurate test results when they are informed of the risk of COVID-19 infection, their enthusiasm for using such applications will be reduced.

The government should take responsibility of testing COVID-19 for the public, providing convenient and affordable testing approaches for them.

\section{Conclusion}\label{sec:conclusion}

With the global pandemic of COVID-19, how to use technology to assist in tracing and suppressing the spread of COVID-19 has become one of the focuses of researchers. This article gives an overview on Bluetooth-based COVID-19 proximity tracing proposals. We categorized the protocols into two categories and summarized the differences between them. Then we specifically analyzed the five protocols and summarized their features and benefits.
For a deeper comprehension, we summarized eight security and privacy design goals of proximity tracing proposals and analyzed the five proposals' achievement of these goals. We found that none of them has achieved the design goals. Moreover, we shed light on the numerous open issues and opportunities that need further research efforts from the technical requirements and community building perspectives.






\bibliographystyle{IEEEtran}
%

\bibliography{refs}

\vspace{-20pt}
\begin{IEEEbiographynophoto}
	{Meng Shen} (M'14) received the B.Eng degree from Shandong University, Jinan, China in 2009, and the Ph.D degree from Tsinghua University, Beijing, China in 2014, both in computer science. Currently he serves in Beijing Institute of Technology, Beijing, China, as an associate professor, the School of Computer Science, Beijing Institute of Technology. His research interests include privacy protection for cloud and IoT, blockchain applications, and encrypted traffic classification. He received the Best Paper Runner-Up Award at IEEE IPCCC 2014. He is a member of the IEEE.
\end{IEEEbiographynophoto}

\vspace{-30pt}

\begin{IEEEbiographynophoto}
	{Yaqian Wei} received the B.Eng degree in computer science from Xidian University, Shanxi, China in 2020.  Currently she is a master student in the School of Computer Science, Beijing Institute of Technology. Her research interest includes cyber security.
\end{IEEEbiographynophoto}

\vspace{-30pt}

\begin{IEEEbiographynophoto}
	{Tong Li} received his B.S. degree from Wuhan University in 2012 and his Ph.D. degree from Tsinghua University in 2017. He is currently a senior researcher in the Computer Network and Protocol Lab at Huawei. His research interest includes network protocols, security and measurements. 
\end{IEEEbiographynophoto}

\end{document}